\documentclass[english,prb,showpacs,amsmath,amssymb,aps,mathbbold,12pt]{revtex4-1}

\usepackage[latin9]{inputenc}
\usepackage[english]{babel}
\usepackage{esint}
\usepackage{braket}

\usepackage{graphicx}
\usepackage{epstopdf} 
\usepackage{pdfsync}
\usepackage[dvipsnames]{xcolor}
\usepackage[normalem]{ulem}
\usepackage{hyperref}
\usepackage{slashed}
\usepackage{dsfont}
\usepackage{tikz}
\usetikzlibrary{snakes}

\usepackage{subcaption}

\newcommand{\be}[1]{ \begin{eqnarray} \mbox{$\label{#1}$} }
   
\newcommand{\ee}{\end{eqnarray}}

\newcommand{\eeq}{\end{equation}}

\newcommand\ie {{\it i.e. }}
\newcommand\eg {{\it e.g. }}
\newcommand\etc{{\it etc. }}
\newcommand\etcp{{\it etc.. }}

\newcommand\half{\frac 1 2 }

\newcommand{\etalc} {{\em et al., }}

\newcommand{\av}[1]{\langle #1\rangle}

\newcommand\oncite [1] {Ref.\, \onlinecite{#1} }
\newcommand\oncites [1] {Refs.\, \onlinecite{#1} }

\newcommand\oncitesp [1] {Refs.\, \onlinecite{#1}. }
\newcommand\oncitec [1] {Ref.\, \onlinecite{#1}, }

\begin{document}
\title{ Mean Field Theories of  Quantum Hall Liquids Justified:\\
Variations on  the Greiter  Wilczek  Theme}
\author{T. H. Hansson}
\affiliation{ Fysikum, Stockholm University, Stockholm, Sweden  }
\author{S. A. Kivelson}
\affiliation{Dept.\ of Physics, Stanford University, Stanford, CA., USA }

\begin{abstract}
We present a field theoretic variant of the Wilczek -- Greiter adiabatic approach to Quantum Hall liquids.  Specifically, we define a Chern-Simons-Maxwell theory such that the flux-attachment mean field theory is exact in a certain limit.  This  permits a systematic way to justify a variety of useful approximate approaches to these problems as constituting the first term in a (still to be developed) systematic expansion about a solvable limit.

\end{abstract}
                                        
\maketitle                 

\section{Introduction }                  

In an article from 1990 Greiter and Wilczek (GW) argued\cite{greiter1} that by an adiabatic evolution one can move from a number of filled Landau levels of electrons, to quantum Hall states in the Jain hierarchy. The idea is to slowly concentrate part of the magnetic flux in thin tubes located at the position of the particles. During this process the constant background magnetic field is diminished (or increased if the flux tubes are directed opposite to the background field), and the intermediate states are naturally interpreted as a system of anyons in an effective B field. Clearly this can be thought of as a concrete way to construct Jain's composite fermions. In a later paper, 
GW gave a more formal argument for their construction using special Hamiltonians with singular interactions for the Laughlin states, and also generalized the idea to include the nonabelian Moore-Read Pfaffian state\cite{greiter2}. 
In a very recent paper, they have also constructed an interpolating Hamiltonian for the $\nu=2/5$ Jain state using a technique that might generalize to the other states in the positive Jain series\cite{GW2021}.

We now give a field theoretic version of a distinct but strategically similar construction based on an adiabatic evolution of states of  composite particles. As in the GW case, one limit is an exact rewriting of the Quantum Hall problem, while the other limit is a solvable point. The difference is that  the intermediate states in the GW formulation are states of anyons in a constant magnetic field, while in our approach the intermediate states are
for bosons or fermions interacting with a mildly fluctuating dynamical gauge field.

Our composite particles are described by the following Ginzburg-Landau-Chern-Simons-Maxwell (GLCSM) Lagrangian density
 \footnote{The same effective action was considered in \oncite{mulligan} some time ago in the context of a study of a clean system close to  a putative quantum critical point at the transition between an isotropic and a nematic quantum Hall fluid. The topologically massive gauge theory was introduced by Templeton \etalc in \oncitec{templeton} and discussed in the context or fractional statistics in Refs. \onlinecite{dualityanyons} and \onlinecite{wilsonloops}.}
\be{GLCSM}
{\cal L}_{GLCSM} &= & {\cal L}_{mat} + {\cal L}_{g}  \\ 
{\cal L}_{mat}&= & \psi^\star (i\partial_0 - a_0 +eA_0) \psi - \frac 1 {2m^\star} |(\vec p -e\vec A + \vec a )\psi |^2 - V[\rho]  \nonumber \\
{\cal L}_{g}&= &\frac 1 {2\pi q} \epsilon^{\mu\nu\sigma} a_\mu \partial_\nu a_\sigma + \frac \varepsilon {2g^2} \vec e \cdot \vec e-\frac 1{2g^2} b^2 , \nonumber
\ee 
where  $V$ is a repulsive potential 
which is a functional of the density, $\rho = \psi^\dagger \psi$ (\ie it may  be non-local in space), $e_i=-\dot a_i - \partial_i a_0$ and $b=\epsilon^{ij} \partial_i a_j$ are, respectively, dynamically fluctuating electric and magnetic fields, and $A_\mu$ is the background electromagnetic gauge potential where $\vec\nabla \times \vec A = B$ and $eA_0(\vec r)$ is a one-body potential that may reflect the presence of disorder, controlled gate potentials, \etcp
\footnote{More properly, ${\mathcal L}_{g}$ should be written in terms of two coupled Chern-Simons gauge fields\cite{fradkinnayak} to avoid difficulties associated with the fractional $1/q$ Chern-Simons level numbers\cite{seiberg}.  Without downplaying the importance of this subtlety, we will finesse it here in the interest of clarity of presentation.}
{Except when explicitly mentioned we shall put $\hbar = c =1$.}
For $q=$ odd we take $\psi$ to be  a complex scalar (bosonic) field corresponding to the composite bosons of MacDonald and Girvin\cite{composite} in the form introduced by Zhang and us\cite{ZHK},  
while for $q=$ even, we take $\psi$ to be a Grassman field representing Jain's composite fermions\cite{jainbook} in the form introduced by Lopez and Fradkin\cite{lopez1991,lopez2004}.  

In the $g \to \infty$ limit,
 $a_0$ is a Lagrange multiplier that  enforces the ``flux attachment'' constraint $ 2\pi\, q\, \rho = b =\epsilon^{ij}\partial_i a_j$ that attaches $q$ flux quanta to each composite particle. In this limit, with $m^\star$ set equal to the bare effective mass of the electron, Eq. \ref{GLCSM} is nominally an exact rewriting of the original problem in terms of new composite particles.

However, for finite $\varepsilon$ and $g$, the flux tubes attached to each particle have finite extent, and so have dynamical consequences beyond encoding the particle statistics.  This action no longer corresponds precisely to the problem of physical interest.  However,  we shall see that the approximate mean-field treatment of  flux attachment, which has been so successfully used to analyze various aspects of this problem, is exact in the limit $g\to 0$. {Thus, this construction allows for an explicit field theoretic rendering of an adiabatic evolution, in the spirit of the 
GW thought experiment, from the soluble small $g$ limit to the physical limit, $g\to \infty$.}


\section{Expression in terms of effective interactions}

There are several ways to look at the problem defined by the Langrangian in Eq. \eqref{GLCSM}.  We can integrate out the fluctuating statistical gauge field perturbatively resulting in a renormalization of the chemical potential (which we leave implicit) and  an effective interaction between particles represented by a contribution to the effective action of the form 
\begin{align}
   \delta S =- \half \int d\vec r_1 d\vec r_2 d\tau_1 d\tau_2 j_\mu(\vec r_1, \tau_1)  \hat D_{\mu,\nu}(\vec r_1-\vec r_2,\tau_1-\tau_2) j_\nu(\vec r_2,\tau_2) + \ldots
\end{align}\\
where $ \hat D$ is the bare photon propagator, and $\ldots$ represents higher order terms. The photon in question is massive (gapped) corresponding to the poles in $D(\vec k, \omega)$, the Fourier transform of $\hat D$, with
\be{disprel}
\omega^2  =   \Omega^2 + \varepsilon k^2
\ee
where $k\equiv|\vec k|$,  $\Omega = \mu/\varepsilon$ is the energy gap, 
and $\mu  = g^2 /2\pi q$  
is the familiar topological mass\cite{templeton}. 
Note that in the $\varepsilon \to 0$ limit, where the gauge field has no independent dynamics,  the gap tends to infinity.

Despite the fact that the photon is massive, it induces long-range statistical interactions between the composite particles that come from the
off-diagonal part that couples current and charge, 

\be{chcurrprop}
D^{i0}(\vec k,\omega) =  2\pi q 
\frac{i\epsilon^{ij} k_j}{k^2} \, G(\vec k, \omega) \, ,
\ee
with 
\begin{align}
    G(\vec k,\omega)= \frac {\mu^2}{\mu^2 + \varepsilon k^2 - (\varepsilon\omega)^2 }\, ,
\end{align}
{and, needless to say, it is this very interaction that endows the quasipartiles, \ie the vortices, with fractional statistics as explained in the classic paper  \oncite{ASW} by Arovas, Wilczek and Schrieffer. }
(The full expression for $D_{\mu\nu}$ is given in Appendix A.)
In the limit $g\to\infty$, where $G(\vec k,\omega) \to 1$, this simply corresponds to attaching $q$ flux quanta to each composite particle, turning them back into the original electrons. 

For our discussion, the most important consequence of including the Maxwell terms in ${\cal L}_{g}$ is that they give a size to the flux tubes bound to the charges. The profile of the statistical magnetic field associated with a static 
composite particle at the origin is
\be{profile}
b(r) =  \frac { q} {
\lambda^2}K_0(r/\lambda)
\ee
where $K_0$ is a modified Bessel function and the scale is given by $
\lambda= \sqrt\varepsilon/\mu = 2\pi q \sqrt\varepsilon/ g^2$. At an intuitive level, in the limit that  
$\lambda$ is large compared to the mean spacing between electrons,  
\ie when 
$\pi\lambda^2\ \bar \rho\  \gg 1$ where $\bar \rho$ is the mean electron density, it should become increasingly possible to replace the fluctuating statistical magnetic field by its mean, $b(\vec r) \to \bar b=2\pi q \bar \rho$.  The actual relation between flux and charge is, of course, retarded, but for small $\varepsilon$, so long as the frequencies characterizing the electron dynamics satisfy the inequality,  $ \omega\ll \Omega$, this should be negligible as well. Typically, the energies of most interest are set by the scale of the Coulomb interaction, $V_c \sim e^2 \sqrt{\bar\rho} \sim e^2/\ell$, where
 $\ell \equiv \sqrt{\phi_0/2\pi  B}$ is the magnetic length, and the second estimate is valid when the filling factor, $\nu\equiv \bar \rho\phi_0/B \sim 1$. 

For completeness, a derivation of  equations \eqref{disprel} to \eqref{profile} are given in Appendix \ref{app:CSM}. We have not attempted to renormalize the full coupled GLCSM theory \eqref{GLCSM} but we do not envision that would qualitatively change our results.


\section{Effective gauge theory} \label{sec:gtheory}
An alternative approach is to integrate out the matter fields, leaving us with a description fully in terms of the fluctuating gauge fields. 
{This can be done perturbatively in powers of the amplitude of the fluctuations of the gauge fields about their saddle point values, with the leading order approximation being a version of RPA theory.  The first step is to integrate the $\psi$-field to get an effective action for the gauge field $a$:
\be{effgauge1}
S^{eff}[a]=S_{mat}[a-eA] + S_g[a]
\ee
and then identify the static saddle point configurations, $\bar a$ for $S^{eff}[a]$
by solving the classical equations of motion. Correspondingly, 
the mean fields, $\tilde B$ and $\tilde E_i$, felt by the composite particles, can be computed from $\tilde A \equiv  eA - \bar a$. The resulting background gauge fields, are related to the expectation values of charges and currents by (see Appendix \ref{app:CSM}).
\be{effgauge}
    \bar b(\vec k) &=& \epsilon^{ij} i k_i \bar a_j =  2\pi q\, G(k,0) \left(\langle \rho(\vec k)\rangle + \frac{\varepsilon}\mu \, \epsilon^{ik} i k_i \langle j_k(\vec k)\rangle \right)  \\
   \bar {e}_i(\vec k) &=& i k_i \bar a_0 -i\omega \bar a_i = 2\pi q\, G(k,0) \left(  \epsilon^{ik} \langle j_k\rangle +\frac{ik_i}\mu \langle \rho(\vec k)\rangle\right)\nonumber \, .
\ee
For example, in the composite fermion representation of an ideal quantum Hall state, $\tilde B$ is such that the composite fermion filling factor, $\tilde \nu \equiv \bar \rho \phi_0/\tilde B$ is an integer, while in the composite boson representation, $\tilde B=0$.
Conversely, for electron filling $\nu=1/2$, the composite fermions with $q=2$ so $\tilde B=0$, while the composite bosons with $q=1$ experience a field corresponding to  $\tilde \nu=1$. In the $g \to \infty$ limit, where $G\to 1$ and $\lambda \to 0$, we regain the usual Chern-Simons expressions,
 $\bar b(\vec k)=(2\pi q) \ \langle \rho(\vec k)\rangle$ and $ {\bar e_i}(\vec k) = (2\pi q)\epsilon^{ik}\ \av{j_k(\vec k) }$;
 on the other hand, in the $g \to 0$ limit, since $G(k,0) \sim (k\lambda)^{-2}$ for $k\lambda \gg 1$, even in the presence of disorder where $\langle \rho(\vec k)\rangle$ and $\langle \vec j(\vec k)\rangle$ are non-zero for non-zero $k$, $\bar b(\vec k)$ and $\vec e(\vec k) \to 0$ for $k\neq 0$, while  $\bar b(\vec 0)= 2\pi q \ \bar \rho$ {and $ \bar e_i(\vec 0) =2\pi q \epsilon^{ik} \langle  j_k(\vec 0) \rangle$. }

The next step is to expand in the fluctuations $\delta a$ around the saddle point $\tilde A$, where for later convenience we also introduce a probe field, $eA \rightarrow eA + e\delta A$, 
in order to identify appropriate response functions.  With this, and  by a shift $\delta a\rightarrow e\delta A + \delta a $, we get

\be{seff}
    S^{eff}[a]&=&\int d\vec r d\tau\ {\cal L}_{g}(\bar a)  +S_{mat}[\tilde A]  + \int d\vec r d\tau\ {\cal L}_{g}(\delta a+e\delta A) \\
    &-&\half\int d\vec r_1 d\vec r_2 d\tau_1 d\tau_2  \ \delta a_\mu(\vec r_1,\tau_1)\ \hat\Pi^{\mu\nu}(\vec r_1,\vec r_2; \tau_1 -\tau_2)\ \delta a_\nu (\vec r_2,\tau_2) + \ldots \nonumber
\ee
where we use the notation $\mu =(i,0)$ but make no distinction between upper and lower indices. 
{We have used a real time formulation where $S_{mat} =  E_{mat}t$ with $E_{mat}$ the ground-state energy of the matter fields in the presence of a static background gauge field, $\tilde A$. In the Euclidean formulation $S_{mat} = F_{mat}/T$ where $F_{mat}$ is the free energy.} $\hat \Pi(\vec r_1, \vec r_2, \tau_1-\tau_2)$ encodes the linear response of the matter fields to an external gauge field, and $\ldots$ represents the non-linear response of the same matter fields. 

In the translationally invariant case, the Fourier transform of $\hat \Pi$, $\Pi^{\mu\nu}(\vec k,\omega)$, can be expressed\cite{dunghai} in terms of more familiar response functions as
\be{seff2}
    \Pi^{ij}(\vec k,\omega) &=& i\omega\tilde \sigma_{ij}(\vec k, \omega) \nonumber \\
    \Pi^{i0}(\vec k, \omega)&=& - \Pi^{0,i}(\vec k, \omega) = -ik_j\tilde \sigma_{ij}(\vec k, \omega) \\
      \Pi^{00}(\vec k, \omega)&=& \tilde\kappa(\vec k, \omega)\, , \nonumber
\ee
where $\tilde \sigma_{ij}$ is the composite particle conductivity tensor  and $\tilde \kappa$ is the density-density response function that
we refer to as the composite particle compressibility since  
it is equal to  the thermodynamic compressibility as $\vec k$ and $\omega$ tend to zero.}

Needless to say, the state of the matter fields depends not only on their statistics and their interactions, but also on the values of the static mean-fields, {$\tilde B$ and $\tilde E_i$}.
However, even in this case, the spatial variations should be less and less pronounced the larger $\lambda$.

Finally, to the extent that the higher order terms in $\delta a$ can be ignored, it is possible to integrate over the statistical gauge fluctuations exactly.  Taking appropriate derivates of the resulting expression with respect to the probe fields, $\delta A$, allows us to express the final results for the electron response functions, in particular the resistivity tensor, $\rho_{ij}$, in terms of the composite particle response functions as
\begin{align}
\label{correspondence}
    &\rho_{xx}(\vec k,\omega) = \tilde \rho_{xx}(\vec k,\omega) + i\omega \varepsilon\left( \frac { q}{\mu} \right) G(k,\omega) + \ldots\\
    &\rho_{xy}(\vec k,\omega) = \tilde \rho_{xy}(\vec k,\omega) +  q G(k,\omega)\nonumber  + \ldots\\
   &\kappa(\vec k,0) = \tilde \kappa(\vec k,0)\left[1 + \tilde \kappa(\vec k)\left(\frac { q} \mu\right) G(k,0)\right]^{-1}  + \ldots  \nonumber
\end{align}
where we  restored $\hbar$ and  expressed the results in units of  the Klitzing constant $h/e^2$.
The $\ldots$ is to remind us that we neglected the effects of higher order terms in $\delta a$.

The self-consistency of this approach -- \ie  ignoring the higher order terms, $\ldots$ -- can be justified only if the the gauge field fluctuations are small, \ie so long as
{
\begin{align}
    \langle[b(\vec k) - \bar b(\vec k)]^2\rangle \ \ll \ \left( 
    \phi_0 \right)^2.
    \label{small}
\end{align}
A more detailed discussion of this condition is given in Appendix \ref{app:fluctuations}.}


\section{Application to gapped states}
The basic idea in the GW approach is to adiabatically connect the difficult FQH problem to a simple solvable one without closing the gap. For Abelian states there are two ways of doing this. In the composite boson approach the 
large $g$ limit corresponds   to hard core composite bosons in zero effective magnetic field, i.e. $\tilde B=0$. 
Here, so long as the disorder is not too strong, the composite bosons can be expected to condense into a superconducting state. Because of the charge flux connection inherited from the Chern-Simons term, the Mei\ss ner effect in the superconductor translates into incompressibility of the Hall liquid,  the zero resistance of the superconductor into the quantized Hall conductance, and flux quantization into the existence of quasi-particles with fractional charge and statistics\cite{ZHK,zhang1992chern}.
In the composite fermion picture, the solvable limit is when the effective magnetic field, $\tilde B =\tilde \nu^{-1}\bar \rho \, \phi_0$ with $\tilde \nu^{-1}$  an integer, $n$, so that (provided the disorder and the interaction strengths are not too large) the composite fermions fill $n$ Landau levels.  This state\cite{jainbook} is  incompressible with a gap that is adiabatically connected to the cyclotron gap, and the other properties of the fractional quantum Hall state can be inferred by applying versions of the arguments that have been applied in the integer case.

In order to connect to the analysis of fluctuations in the previous section, we shall here use the composite boson approach, but also comment on the composite fermion picture.

\subsection{Abelian states }

\begin{figure}
\begin{subfigure}{.5\textwidth}
  \centering
  \includegraphics[width=.8\linewidth]{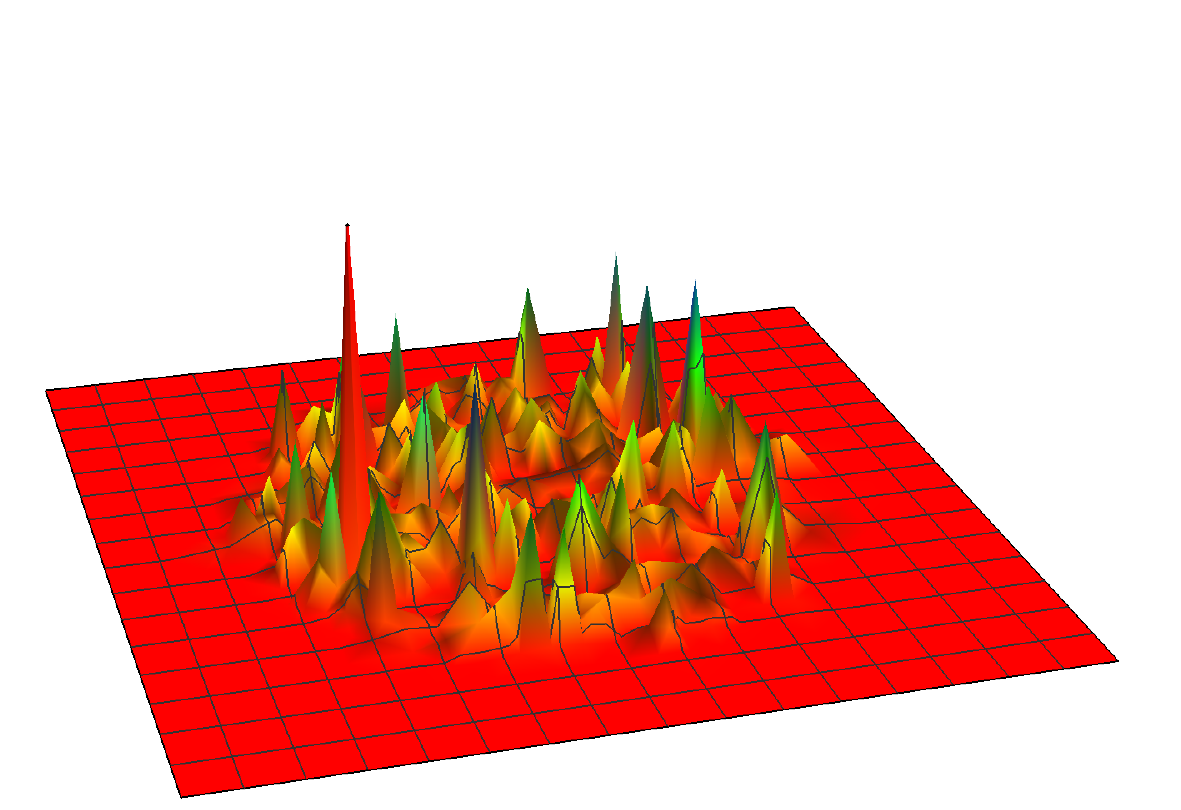}
  \caption{$\lambda = \ell/2$}
  \label{fig:sfig1a}
\end{subfigure}%
\begin{subfigure}{.5\textwidth}
  \centering
  \includegraphics[width=.8\linewidth]{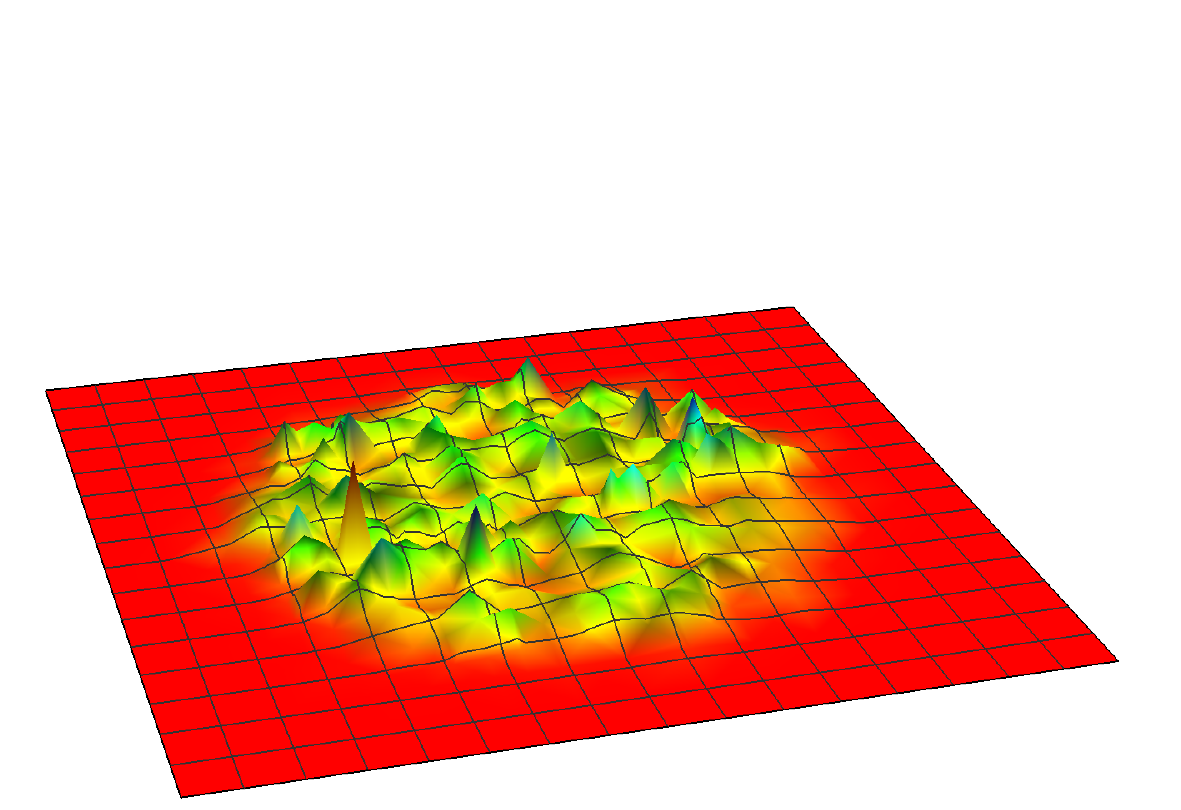}
  \caption{$\lambda = \ell$}
  \label{fig:sfig1b}
\end{subfigure}

\begin{subfigure}{.45\textwidth}
  \includegraphics[width=.8\linewidth]{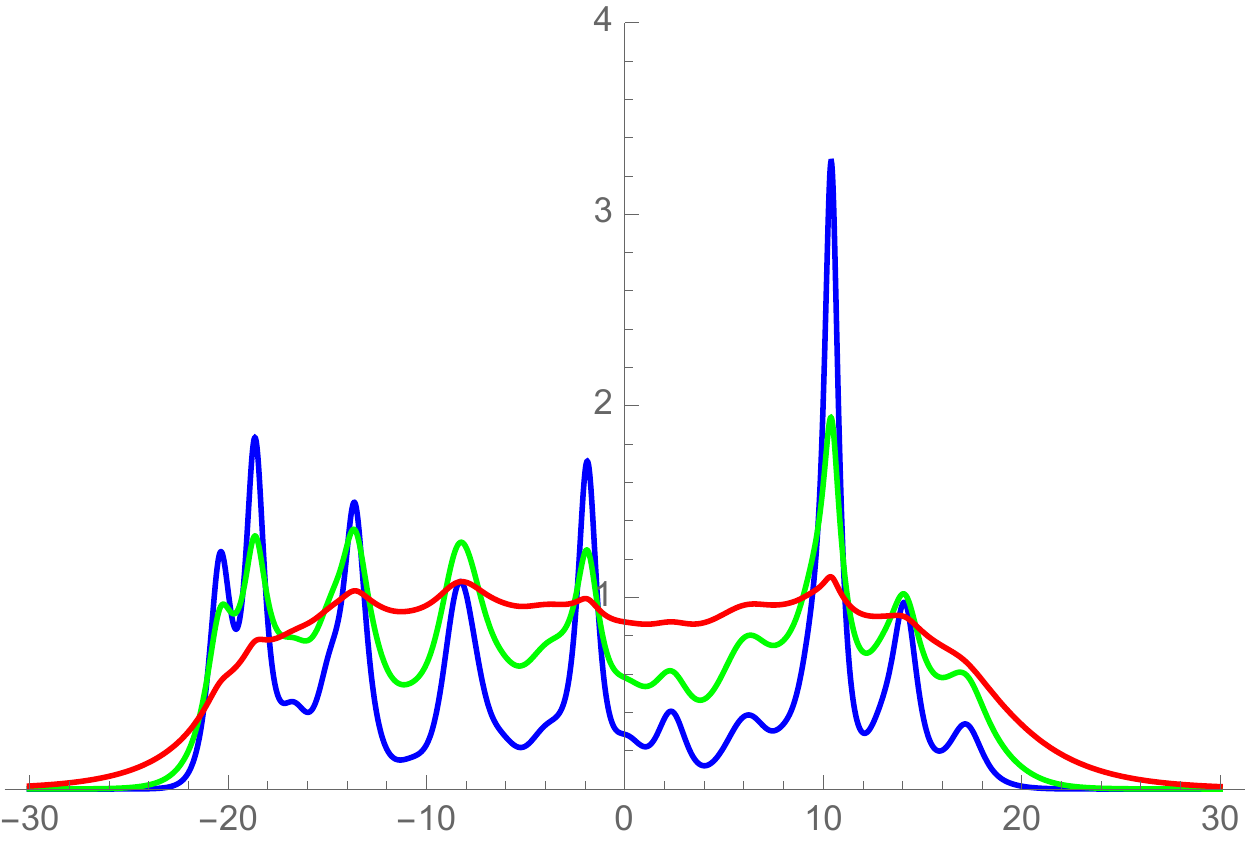}
  \caption{ Cut along x-axis }
  \label{fig:sfig1c}
\end{subfigure}
\begin{subfigure}{.4\textwidth}
  \includegraphics[width=.8\linewidth]{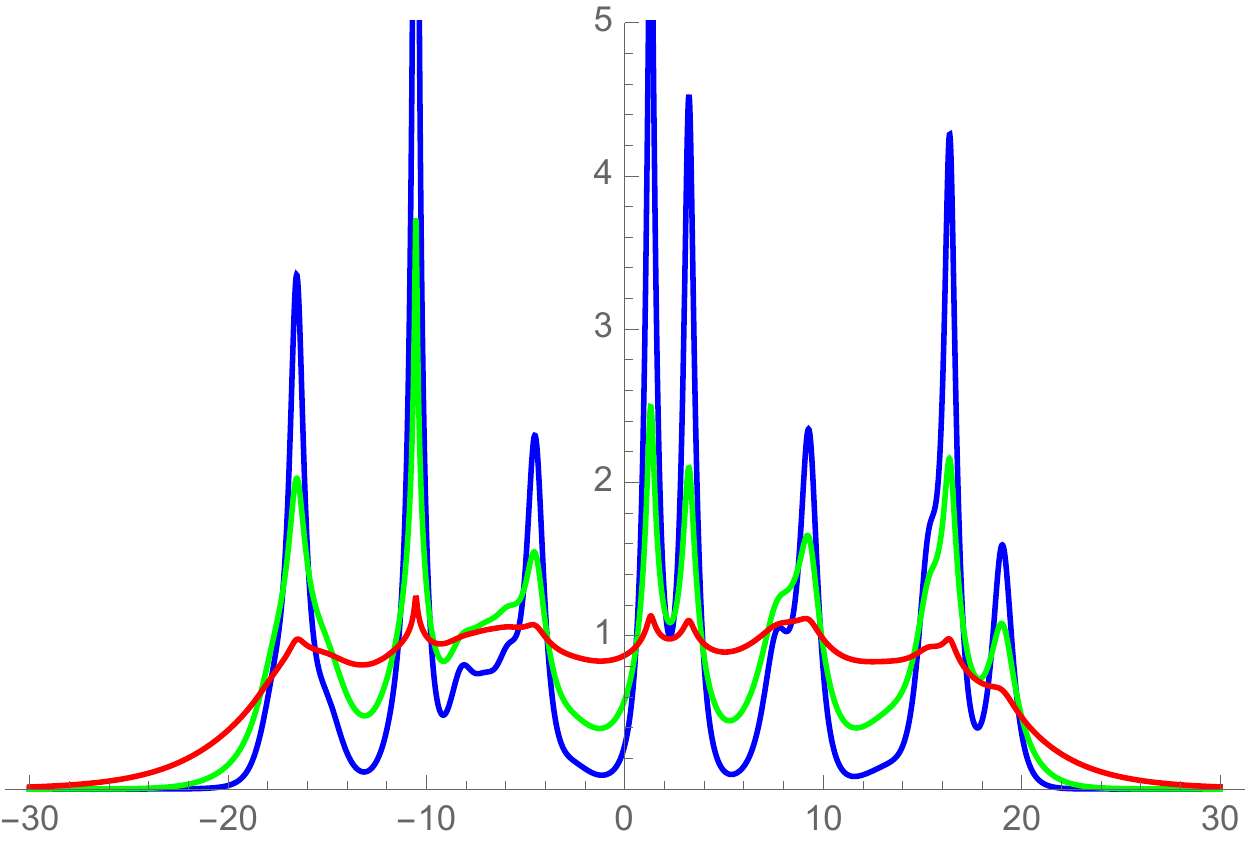}
  \caption{Cut along y-axis  }
  \label{fig:sfig1d}
\end{subfigure}
\caption{ Contour plots of the
statistical $b$-field configurations for a typical snapshot of 200 electrons in a circular $\nu=1$ QH state for (a) $\lambda=\ell/2$ and (b) $\lambda = \ell$, where $\ell$ is the magnetic length. (c) and (d) show the $b$ field along two perpendicular diameters of the droplet. The blue, green and red curves correspond to $\lambda/\ell = 0.5$, 1, and 2 respectively.}
\label{fig:fig1}
\end{figure}

The Laughlin states provide the simplest case where an adiabatic 
evolution from $\lambda \gg \ell$ to $\lambda \to 0$ takes us from a system in which the mean-field approximation is justified to the physical problem of interest.   If  this can done without closing the gap, all the 
``essential'' properties derived from the mean field theory should also hold for the exact theory. We have no proof, but we believe that it is a very likely scenario. Moreover, because we are dealing with incompressible states, we expect the mean-field treatment to be self consistent, and highly accurate at least for the  range of $g$ where $\pi \lambda^2\bar \rho>1$.
In Fig. \ref{fig:fig1} we show the  field, $b(\vec r)$, corresponding to a typical configuration of particles (chosen using the square of a $\nu=1$ Quantum Hall wave-function as a Boltzman weight) for a few values of $\lambda$;  one can see that $b(\vec r)$ rapidly approaches a configuration independent constant with increasing $\lambda$.

This argument for the Laughlin states also applies, \emph{mutatis mutandis}, to the Halperin $(n,n,m)$ bilayer states\cite{halperin1983theory,girvin1995multi} if we properly take into account that in this case we need two statistical gauge fields\cite{blokwen90}.  This will affect details of our estimates of the fluctuations, but will not alter them qualitatively.  

The Laughlin states can be viewed as the subset of the Jain states with one filled Landau level. ``Hierarchy'' states with more than one filled Landau level of composite fermions are most easily understood in the Fradkin-Lopez\cite{lopez1991} field theory of composite fermion: by attaching $\pm 2p$ flux quanta to each particle and again decreasing $g$ until the mean field approximation is valid.  In this case, one finds 
incompressible states corresponding to $n$ filled Landau levels at filling factors $\nu = n/(2np\pm 1)$. The bosonic picture for these hierarchy states is more subtle. Superficially they look very similar to multi-component states. For example, the Halperin (3,3,2) has $\nu=2/5$ just as the leading Jain state. The difference is that in the Jain states there is only one type of 
composite fermion, so the state has to be fully anti-symmetric. In the composite fermion approach this is possible because the electrons reside in two different ``effective Landau levels." The physical meaning of this is that the electrons carry different orbital spins\cite{wen1995topological}, and this must be incorporated in the composite boson picture. How to do this was shown in a recent paper\cite{10.21468/SciPostPhys.8.5.079}, which however also relied on a mean filed approximation. Combining those methods with the adiabatic approach of this paper provides a field theoretic understanding not only of the Jain states, but of a larger class of hierarchy states.

{From \eqref{effgauge} we can also obtain the resistivity tensor at finite $\vec k$.  Specifically, when composite bosons are condensed in a superfluid state, the screening of the magnetic and electric fields persists to finite $\vec k$, so that in equilibrium, and with a $\lambda$ large enough to justify the mean-field approximation, $\tilde B(\vec k)=0$ and $ {\tilde E_i}(\vec k)=\vec 0$. At finite $\vec k$, $\tilde B = \bar b = 0$, which allows us to compute $\vec E$ in terms of $\vec j$ from 
 \eqref{effgauge}, and thus to extract (see Appendix \ref{app:derivation})  the conductivity tensor which we express as, 
\be{finitek}
\frac {\sigma_{xy}(\vec k)} {\sigma_{xy} (\vec 0)}= 1 + \frac{(1-\varepsilon)}{\varepsilon} \left(  k \lambda \right)^2 + O(k^4)
\ee
{This can be compared whith the expression derived  by Hoyos and Son using Galilean invarianace\cite{hoyosson} 
\be{hoyosson}
\frac {\sigma_{xy} (\vec k)}{\sigma_{xy} (\vec 0)}= 1 + \frac s 2 (k\ell)^2+ O(1/\omega_c)
\ee
where $s$ is the orbital spin of the composite particles. 
At a phenomenological level we can match these expressions by adjusting $\epsilon$ and $\lambda$, but within our framework there is no reason for doing so. }

In this context we note that the flux tubes associated with the electrons do carry angular momentum. The field angular momentum is }
\be{fieldL}
L_g = \frac 1 {g^2} \int d\vec r\,  b\, (\vec r\cdot \vec e )
\ee
and using $\vec e = (1/\mu) \vec\nabla b$ that follows form \eqref{effgauge} combined with the expression \eqref{profile} for the $b$-field associated to a point charge, the integration can be done analytically and gives,
\be{orbspin}
L_g = \frac 1 \varepsilon\, \frac q 2 \, .
\ee 
{independent of $\lambda$. This means that the total angular momentum is not changed as we interpolate between a thin and a thick flux tube.
In the Lorentz invariant case, $\varepsilon = 1$, we reproduce the expected value for the orbital spin (the thin vortex limit was given in  Ref. \onlinecite{hagen85}). We note that in this case the $\vec k$ dependence of $\sigma^{xx}$ vanish, but this might be purely accidental. 
}

\subsection{Nonabelian Pfaffian states}

In their second paper\cite{greiter2}, 
GW also applied their adiabatic approach to a state with a filled Landau level of bosons, 
\be{boseinit}
\Psi_{MR}  = \mathrm{Pf} \left( \frac 1 {z_i - z_j} \right) \prod _{i<j} (z_i - z_j)\ ,
\ee
a state that requires a strong repulsion to be stabilized. By adiabatically concentrating half of the flux to the particles they ended up with a state at $\nu=1/2$, and a wave function only differing from \eqref{boseinit} by an extra Jastrow factor $\prod _{i<j} (z_i - z_j)$ which is precisely the nonabelian Moore-Read Pfaffian state\cite{MOORE1991362}. 

In two later papers\cite{gww1,gww2} with Wen, GW instead started from fermions in zero magnetic field, and argued that adding a small CS-term would trigger a superconducting instability giving rise to a $p_x + ip_y$ paired state.  The connection between the QH system and p-wave paired (spinless) superconductors was later discussed in detail by Read and Green\cite{readgreen}.

We shall connect to this latter approach, \ie we 
consider the adiabatic continuation from an \emph{assumed}  superconducting state of spinless fermions in zero magnetic field, to the MR Pfaffian states of composite fermions in a background field at $\nu=1/2$.  {Clearly the existence of such a superconducting state is contingent upon the details of the interaction. Also, superconductivity is not the only possible state in the small $\lambda$ limit, and in the next section we shall discuss the other obvious candidate, namely the Fermi liquid. }

Much of the interest in the Pfaffian state 
derives from its nonabelian quasiparticles.  The paired superconductor hosts vortices that binds Majorana states which are at the origin of the nonabelian statistics\cite{ivanov}. If one could verify that these vortices are present also in the full GLCSM theory, our construction would give a strong argument for having a nonabelian QH state at $\nu=1/2$ that is not based on an explicit wave function, but rather on a well justified mean field theory.


\section{Compressible states}
There are at least two circumstances in which compressible states are thought to arise in quantum Hall systems:  1) The first is a composite fermion metal that occurs when  the net statistical flux exactly cancels the total magnetic flux, \ie when
$\tilde B$ for the composite fermions vanishes.  Such a state might exist either in the absence of disorder or in the presence of weak disorder.  2) The other is the critical state that occurs at a continuous transition between two quantum Hall plateaus or at the transition between a quantum Hall state and an insulator. Since, in the absence of disorder\cite{firstorder}, these transitions are generically first order, this state presumably exists only in the presence of suitable disorder.  Clearly, for either of  these states, the notion of adiabatic continuity is  more subtle than for inccompressible states.  Specifically, even though the mean field treatment is exact in the limit $\lambda \to \infty$, it is a non-trivial issue to what extent the results continue smoothly to the case of large but finite $\lambda$. 


\subsection{The Composite Fermion Metal}
The effect of dynamically fluctuating gauge fields on the  composite fermion metal is a complex problem\cite{simonandbert,RevModPhys.75.1101}. 
Typically, the Fermi liquid state is found to be perturbatively unstable, so various clever tricks - including an early and highly influential approach by Nayak and Wilczek\cite{nayak} - have been adopted to address it. It is a potentially revarding challenge to devise a systematic approach that exploits the twin small parameters, $\lambda^{-1}$ and $\varepsilon$.

However, in the presence of disorder, there is an interesting piece of physics that can be extracted for large but finite $\lambda$.  To be explicit, consider the case of most direct experimental relevance, that is $\nu=1/2$, which for composite fermions implies  $q=2$.  Now, for $\lambda=\infty$, $\tilde B=0$, which is to say that the composite fermions see no trace of the external magnetic field.  Consequently, the composite fermion Hall response vanishes, $\tilde\sigma_{xy}=0$.  However, this result cannot be generically correct since the electrons are coupled to the magnetic field, and indeed it is possible to prove\cite{meanddunghai,shankarparticlehole} that, in the special case where the electronic problem is particle-hole symmetric,  the composite fermion $\sigma_{xy} = -1/2$.  
This apparent contradiction can be resolved in an elegant fashion as  was shown in \oncitesp{sriandmike1,sriandmike2} 
Clearly, in the presence of disorder, the electron density is expected to be inhomogeneous, and for weak disorder, $U$, and small $k$, this can be captured in linear response:
\begin{align}
    \langle \rho(\vec k)\rangle = \kappa(k,0) U(\vec k).
\end{align}
Correspondingly, there will be a spatially varying effective magnetic field, 
as in \eqref{effgauge}.  It is still the case that the average field vanishes, but because the average field is correlated with the particle density, the average of the cross-correlation between field and density is finite,
\begin{align}
    \overline{ \langle \rho^\star(\vec k)\tilde B(\vec k)\rangle} = -2\pi q \ G(k,0) \ \left|\kappa(k,0)\right|^2  \overline{\left|U(\vec k)\right|^2}
\end{align}
where $\overline{O}$ is the average over disorder configurations of $O$.

Indeed, in  \oncites{sriandmike2,sriandmike1}, evidence was presented that this correlation results in an emergent particle-hole symmetry, \ie  it implies $\sigma_{xy}=-1/2$, a conclusion that was reached earlier in Ref. \onlinecite{chongandhalperin} using a complementary approach.  To this we add the control of mean-field theory and the suppression of large $k$ fluctuations of $\tilde B$ that sharpens the analysis.


\subsection{Critical States}

A QH state near a plateau transition becomes compressible.  In the composite boson language, this corresponds to a superconductor  
to insulator transition (SIT), as originally discussed by  Lee, Zhang and one of us\cite{KLZ}. In experiments this transition is always studied as a function of magnetic field at fixed disorder, which in composite boson language corresponds to studying the SIT as a function of increasing $\tilde B$.  Here the critical value $ \tilde B_c$ is related to the critical magnetic field, $B_c$ as $B_c =q\phi_0\bar \rho + \tilde B_c$.  

The important point is that -- to the extent that the RPA description is valid, \ie that
higher order terms in $\delta a$ can be ignored -- all the critical properties of the system are those of  bosons undergoing a field-driven SIT, regardless of the value of $q$.  The implication is that  the critical properties of the system are fundamentally the same for a system undergoing a field driven SIT, a $\nu=1$ to insulator transition (which corresponds to the case $q=1$), \ie an integer plateau transition, or a $\nu=1/3$ to insulator transition (which corresponds to the case $q=3$), \ie a fractional quantum Hall transition.
The authors of \oncite{KLZ} referred to this as a ``law of corresponding states", which has since\cite{dolan} been referred to as ``superuniversality".

To be  more explicit,  consider what we know of the properties of bosons undergoing a field-driven SIT.   For $\tilde B < \tilde B_c$, the system is superconducting, which means that  the composite boson response functions at small $\omega$ and $T= 0$ are $\tilde \rho_{xx}(\vec 0,\omega) = (m^\star/e^2\tilde n_s) i\omega$ and $\tilde \rho_{xy}(\vec 0,\omega)=0$ where $\tilde n_s$ is the superfluid density.
\footnote{A more appropriate description is that $\tilde n_s/m^\star$ is the superfluid stiffness.}
This holds even in the state near  the transition so long as all field-induced vortices (quasi-particles) are localized.
Because these vortices can always diffuse at finite $T$  by variable-range-hopping, there can be no finite temperature transition.  {However,  we expect there to be a characteristic temperature scale, $T^\star$, below which $\tilde \rho_{xx}(\vec 0,0)$ decreases in some exponential manner with decreasing $T$. On the basis of general scaling considerations\cite{sondhigirvin}, one expects that both $T^\star$
and $\tilde n_s$ vanish upon approach to the critical field in proportion to $(\tilde B_c - \tilde B)^{\nu z}$ where $\nu$ and  $z$ are, respectively, the correlation length and dynamical exponent of the  field driven SIT.} 

As can seen from Eqs. \ref{correspondence}, this behavior of the composite bosons can be directly translated into statements concerning the electron response functions.  At $T\to 0$, (ignoring terms of order $\omega^2$ and higher) 
\begin{align}
   & \rho_{xx}(\vec 0,\omega) = i\omega\left(m^\star/e^2 n_s\right) \\
  & \rho_{xy}(\vec 0,\omega) = q, \nonumber
  \label{superfluid}
\end{align}
\ie no matter how weakly superconducting the composite bosons, the corresponding Hall resistance is quantized and $\rho_{xx}\to 0$ as $\omega \to 0$.  
Interestingly, the scaling relations imply that   the reactive portion of $\rho_{xx}$, which is proportional to $1/n_s \equiv 1/\tilde n_s + q \varepsilon e^2/m^\star\mu$, {should diverge in proportion to $(\tilde B - \tilde B_c)^{-\nu z}$ }-- a prediction that, as far as we know, has never been tested.\\

There are general scaling arguments\cite{fisher,KLZ} that suggest that at the critical point of the SIT, the resistivity tensor takes on a universal value, $\tilde\rho^{(c)}_{ij}$.  Again from Eq. \ref{correspondence}, this implies universal values for the resistivity tensor at the point of a quantum Hall transition,
\begin{align}
    &\rho_{xx}^{(c)}(\vec 0,0) = \tilde\rho^{(c)}_{xx} \\
    &\rho_{xy}^{(c)}(\vec 0,0) =  q+ \tilde\rho^{(c)}_{xy} \nonumber
\end{align}
Arguments based on a speculated self-duality at the transition\cite{fisher,KLZ,shaharsondhi,kapitulnik}, suggest that $\tilde\rho^{(c)}_{xx}  =1$ and $\tilde\rho^{(c)}_{xy}=0$.  This is consistent with a distinct set of theoretical expectations in the context of the $\nu=1$ to insulator   transition where by particle-hole symmetry one expects $\sigma^{(c)}_{xy} = 1/2$ and where various arguments\cite{bhatt,tsuibhatt} suggest that $\sigma^{(c)}_{xx} =1/2$ as well -- \ie that $\rho^{(c)}_{xx}=\rho^{(c)}_{xy}=1$.

A similar analysis can be carried out on the insulating side of the SIT.  The results are analogous to the above.

More generally, much of this analysis parallels that of Ref. \onlinecite{KLZ}.
What the present work adds is to identify a limit,  $\lambda \gg \ell$, in which the neglect of the higher order terms in $\delta a$ is justified. 
Of course, there is no a priori reason to think that no corrections to the critical properties  arise at order $1/\lambda$, and these could well depend on $q$ - thus spoiling the superuniversal aspects of the results.  However, it may give some way to understand that - in several experimental circumstances\cite{jiang,tsuibhatt,shaharsondhi} - superuniversality seems to be approximately realized, both with regards to measured critical exponents and the value of the resistivity tensor at criticality.  Indeed, inverting the logic of superuniversality, it is possible to translate known results for the quantum Hall plateau transition into predictions for the properties of the SIT.  Interestingly, recent experiments\cite{kapitulnik} on the SIT in superconducting films have found results -- including apparent particle-vortex self-duality -- that are consistent with the corresponding observations in quantum Hall systems.


\section{Future Directions}
Until now, we have used the Maxwell terms as a tool to adiabatically connect QH states to other known states such as 
superconductors or Fermi liquids. Alternatively we could imagine using $\mathcal{L}_{GLCSM}$
as a phenomenological theory, where the parameters $g$ and $\varepsilon$ could be adjusted to fit \eg the collective Girvin-Platzman and Kohn modes. At a much more ambitious level, it is also possible to imagine that $\mathcal{L}_{GLCSM}$ could emerge as an effective field theory once some high-energy degrees of freedom have been integrated out. It is certainly plausible that an effective size for the flux attachment, $\lambda \sim \ell$, would emerge when states in higher Landau levels are integrated out.

It is plausible, but not inevitable, that the resulting low energy effective theory will still have the same $U(1)$ gauge structure.  It is, after all, the natural theoretical framework for  topological effects like fractional charge and statistics (and possibly also Hall viscosity or shift) which should not be affected by short distance cutoffs, and we would thus expect the crucial current-charge propagator $D_{i0}$ in  \eqref{chcurrprop} to remain the same albeit with a renormalized pole position. Just as in the case with massless photons, where a momentum cutoff typically introduces a mass and thus a new (longitudinal) degree of freedom, this could happen also in our case but such a mode is expected to be massive and thus not destroy the infrared behavior. So, without being explicit, we may assume that we can integrate out high energy modes in a way that preserves gauge invariance with respect to the statistical gauge fields. In this sense, $L_{GLCSM}$ with finite $g$ (or  a  generalized version of it) can be considered to be a course grained version of the ``microscopic'' problem.

If we accept $L_{GLCSM}$ as physical - \ie not just a useful prop in  a thought expepriment - we can use some of the above results to make statements of relevance to the real world.  Since, in the limit $\lambda \gg \ell$, the line of analysis provides a justification for a law of corresponding states relating the behavior of integer and fractional quantum Hall plateau transitions to each other and to the magnetic field driven 
SIT, it is reasonable to conclude that it should apply in an approximate sense so long as $\lambda \gtrsim \ell$.  Possibly this rationalizes {the considerable emperical support for the notion of corresponding states.
One new result of the analysis in this paper }that warrants experimental consideration is the relation given in Eq. \ref{superfluid} that relates the reactive portion of $\rho_{xx}$ at low frequencies to the superfluid density of the composite bosons.  Recall that for a Gallilean invariant system, Kohn's theorem implies that $n_s=\bar \rho$ in this relation.  Exploring whether, and in what  way,   $n_s$ vanishes upon approach to a quantum Hall to insulator transition could reveal a central feature of the SIT of the composite bosons.

We have not touched upon the nature of the quasiparticles. In the bosonic description they are vortex solutions, while in the composite fermionic description they are holes in the various effective Landau levels. In the latter picture the quasiparticles are naturally endowed with an orbital spin, but it remains to understand this in the bosonic version. As hinted at  earlier, we  might speculate that for the extended fluxtubes there might be a more general flux attachment procedure that allows for different orbital spins.


\section{TRIBUTE}

Since this paper is dedicated to Frank Wilczek on his 70-ies birthday, we end with two personal notes.

{ SK: I met Frank when he took up his position at the ITP at UCSB in 1980. 
I was a freshly minted ``solid state physicist" for whom encountering the notion of a unified field of ``theoretical physics" - a notion embodied by the breadth of Frank's interests and contributions - was exhilarating and formative.
I remember telling Frank about work I was completing in collaboration with my mentor, J. R. Schrieffer, showing that  fractional charge (in this case of 1D solitons) can be a sharp quantum observable - something that was much disputed at the time by theorists with frightening credentials.  That Frank found the line of reasoning convincing (although, admittedly, he also thought the answer to be so obvious that it was not worth disputing) was one of those little things that had a great impact on my scientific life.  Years later, in less happy circumstances, a famous quote of Frank's was a source of some comfort: ``If you don't make mistakes, you're not working on hard enough problems.''  At least I have made mistakes.  In recent years - through the catalyst of the friendship between my wife, Pamela Davis, and Frank's charismatic partner, Betsy Devine - I have experienced the joy of inclusion in the Wilczek circle of friends.} 

{THH: I also first met Frank at the ITP in the the fall of 1981, when he and my mentor, Ken Johnson from MIT, co-organized a QCD workshop, but it was only when Frank took up a position in Stockholm in 2016 that I got to know him more closely. I have since witnessed his great impact on the theoretical physics environment at Stockholm University and at Nordita, where he has built a thriving research group. He has also  been instrumental in attracting large grants to expand the activities at Nordita in new and exciting directions. Personally I have had the great privilege to work and discuss with Frank both on physics projects, and on environment building activities at Nordita and at the TDLI. As a regular participant in his group meetings, I have also seen how he mentors and encourages young physicists by creating a challenging and critical, but always kind and supportive atmosphere. Work aside, I have fond memories of the many joyful moments I have spent  with Frank and Betsy, in the gardens of H\"ogberga, in the rollercoasters of Gr\"ona Lund, and at dinners in Stockholm and Shanghai. }

{\bf Acknowledgements:}  
We have got very helpful feedback on this work from several colleagues, and we especially want to thank Srinivas Raghu for numerous discussions, and Mikael Fremling for providing input data for the figures. 
S.A.K. was supported, in part, by NSF grant No. DMR-2000987 at Stanford.


\appendix
\section {Details on the CSM theory} \label{app:CSM}
\subsection{Dispersion relation and fluxtube profile} \label{app:A1}
Using the Coloumb gauge $\vec\nabla\cdot \vec a = 0$ to write $a_i = \epsilon^{ij}\partial_i\chi$, and with a polar representation of the bosonic field,  $\psi = \sqrt\rho e^{i\theta}$, we can write the gauge Lagrangian to quadratic order as,
\be{matrix}
\mathcal{L}_{g} &=& \frac 1 {2\pi q} a_0 k^2 \chi + \frac \varepsilon {4\pi q \mu} a_0 k^2  a_0 - \frac 1 {4\pi \mu} \chi \left[ (k^2)^2 - \varepsilon k^2\omega^2\right]\chi  \\
 &=& \frac 1 {4\pi q \mu} \left( \begin{array}{cc} \chi , & a_0  \end{array} \right)
\left(
\begin{array}{cc}
 \varepsilon  k^2 \omega ^2-k^4 & \mu  k^2 \\
 \mu  k^2 & \varepsilon  k^2 \\
\end{array}
\right)
 \left(
\begin{array}{cc} 
     \chi  \\
     a_0
\end{array}\right) \nonumber \, ,
\ee
where $k^2 = -\nabla^2$, $\omega^2 = -\partial_t^2$ and we recall that  $\mu =g^2/2\pi q$ .
Here and in the following we use the simplified notation $\xi(-\vec k) f(\vec k) \xi(\vec k) = \xi f(\vec k) \xi$ \etcp

To extract correlation functions and response, we introduce a source term 
\be{source}
{\mathcal L}_{source} = - a_0 \rho + \vec a \cdot \vec j = - a_0 \rho +  \chi j_\chi
\ee
where $ j_\chi = i \epsilon^{ij} k_ij_j$. Integrating the gauge field gives the response action
\be{response}
S(\rho,\chi) = \half  \begin{pmatrix} j_\chi , & \rho \end{pmatrix}
\begin{pmatrix} \varepsilon & -\mu \\ -\mu & \varepsilon \omega^2 - k^2 \end{pmatrix}
\begin{pmatrix} j_\chi   \\ \rho \end{pmatrix}   \frac {2\pi q } { \mu k^2} \, G(\vec k,\omega)
\ee
where,
\be{defg}
G(\vec k,\omega) = \frac {\mu^2} {(\varepsilon\omega)^2 - \mu^2 -\varepsilon k^2} \, .
\ee
 Using $b = \nabla^2\rho$ the  fluxtube profile is directly extracted from the $(\chi,\rho)$ component  of the response action,
\be{bprof}
b(k) = k^2 \chi = \frac{{2\pi q \mu^2}/ \varepsilon }{k^2  +\mu^2/\varepsilon} \, \rho = \frac{2\pi q} {(k\lambda)^2 + 1}\, \rho
\ee
which after Fourier transformation gives \eqref{profile} in the main text. 


\subsection{Response functions} \label{app:B1}
 In the translationally invariant case the  part of \eqref{seff} relevant for calculating linear response can be written as,
\be{seffapp}
    S^{resp}[\delta a,\delta A] = \int \frac {d\vec k d\omega}{(2\pi)^4} \left( \half (\delta a_\mu + e\delta A_\mu)\Pi_g^{\mu\nu}  (\delta a_\nu + e\delta A_\nu) - \half \delta a_\mu \Pi_{mat}^{\mu\nu} \delta a_\nu \right)
\ee
and after integrating over the fluctuation $\delta a$ of the statistical field the electronic response function becomes,
\be{elresp}
 S^{resp}_{el}[\delta A] = \int \frac {d\vec k d\omega}{(2\pi)^4} \,\delta A_\mu \Pi_{mat}^{\mu\nu} \delta A_\nu \, .
 \ee
 where $\Pi_{el}$ is expressed in terms of $\Pi_g$ and $\Pi_{mat}$. This relation is simpler when written in terms of the inverse matrices $D_{el} = \Pi_{el}^{-1}$ \etc {(which differ from the standard Minkowski propagator 
 by a factor $i$ )},
 \be{rmat}
 D_{el}^{\mu\nu}=  D_{mat}^{\mu\nu} - D_{g}^{\mu\nu}
 \ee
which are related to the resistivities. 

The most direct way to get the explicit expressions for $D_{g}^{\mu\nu}$, is to write \eqref{matrix} in the $(a_x,a_y,a_0)$ basis,
\be{matrix2}
\mathcal{L}_{g} = \half \left( \begin{array}{ccc} a_x,a_y , & a_0  \end{array} \right)
\frac 1 {2\pi q \mu} \left(
\begin{array}{ccc}
 \varepsilon   \omega ^2-k^2 & 0 & -i \mu k_y \\
0 & \varepsilon   \omega ^2-k^2  &  i \mu k_x \\
i\mu k_y & -i \mu k_x & \varepsilon k^2
\end{array} 
\right)
 \left(
\begin{array}{ccc} 
     a_x  \\
     a_y \\
     a_0
\end{array}\right) \nonumber \, ,
\ee
and then diagonalize this matrix to get,
\be{rmuny}
D^{ij} &=& -2\pi q  \, \frac { \varepsilon} \mu\,  G(\vec k,\omega)\, \delta^{ij} \nonumber \\
D^{i0} &=& - D^{i0} = -2\pi q   \, G(\vec k,\omega) \epsilon^{ij} \frac {i k_j} {k^2} \\
D^{00} &=& -2\pi q  \, \frac{\varepsilon\omega^2 - k^2} {\mu  k^2 } \, G(\vec k,\omega) \nonumber
\ee

The compressibility can directly be read from $D^{00}$ or equivalently from 
\eqref{response}. To extract the resistivity tensor $\rho^{ij} = \rho_{xx} \delta^{ij} + \rho_{xy} \epsilon ^{ij}$, we write the electric field as 
\be{elfield}
e_i = i\omega a_i - i k_i a_0 = i\omega (D^{ij} j_j +D^{i0}\rho) - ik_i D^{0l}j_l
\ee
and then substitute the expressions \eqref{rmuny} and use current conservation, $\dot\rho = \vec\nabla \cdot \vec j$ to get \eqref{correspondence} in the main text. Note that the term $i\omega a_i$ must be included to get the correct  $\omega =0$ result, since there is a term $i\omega\rho = - i\vec k\cdot\vec j$.


\subsection{Derivation of Eq. \eqref{finitek} } \label{app:derivation}
{
Putting the first line in \eqref{effgauge} to zero yields, 
\be{density}
\av\rho = \frac \varepsilon \mu \epsilon^{ik} i k_i \av{j_k}
\ee
and substituting this in the second line of \eqref{effgauge} we get,
\be{statefield}
eE_i = \bar e_i = 2\pi q \frac 1 {1 + \left( \frac k \mu^2\right)} \left[1 + \varepsilon \left(\frac k \mu \right)^2 \right] \epsilon^{ik} \av{j_k} \, .
\ee 
From this we can extract the conductivity,
\be{sonrel}
\frac 1 {2\pi q} \sigma_{xy}(\vec k) =   1 + (1-\varepsilon) \left( \frac k \mu \right)^2 + O(k^4) \, .
\ee
}
Expressing $\mu$ in $\lambda$ we get \eqref{finitek} in the text.

\section{What does it mean for gauge field fluctuations to be small?} \label{app:fluctuations}
In order for the mean-field treatment to be accurate, the fluctuations of the statistical gauge fields must be small.  What this means is most readily addressed from the composite boson perspective.

{To begin with, let us examine what this means for the static fields produced as a response to the (assumed weak) disorder in the system.} If the flux of $\bar b$ through any area of a superconductor exceeds a flux quantum, the energy can be lowered by nucleating a vortex, which constitutes a strong and highly non-linear response.  Conversely, if the magnetic fields are sufficiently small that the flux through any area is small compared to $\phi_0$, the response of the composite bosons is correspondingly small and linear.  In Fourier space, this leads to a condition of the form of Eq. \ref{small}.

Given a particular realization of the disorder potential $U_{\vec k}$, 
the variations in the composite particle density  and current density can be computed in linear response as
\begin{align}
   & \langle \rho(\vec k)  \rangle =  \tilde \kappa(\vec k,0) U(\vec k) + \ldots \, , \\
    &\langle j_i(\vec k)  \rangle =  -i \tilde \sigma_{ij}(\vec k,0) k_jU(\vec k)  + \ldots \, .\nonumber
\end{align}
so, when averaged over impurity configurations, the mean-squared variation of $\bar b$ follows from \eqref{effgauge},
\be{Delta}
\Delta(\vec k) \equiv \overline{\left| \langle b_{\vec k} \rangle \right|^2 } = \left |2\pi q G(k,0)\tilde K(k,0)\right|^2 \ \ \overline{\left |  U_{\vec k}\right|^2}
\ee 
where $\overline{X}$ is the configuration average of $X$  and
{
\begin{align}
    \tilde K(k,\omega) = \tilde \kappa(k,\omega) - \sqrt{\varepsilon}\lambda k_i\epsilon_{ij}\tilde \sigma_{jk}(\vec k,\omega) k_k\, .
\end{align} }

In the same way, we can compute the mean squared magnitude  of the fluctuations of the statistical magnetic field, \\
\be{delta}
\delta(\vec k, \omega) \equiv \langle\, | b_{\vec k,\omega}- \langle b_{\vec k,\omega}\rangle |^2 \rangle =   | 2\pi q G( k,\omega)|^2 \tilde K( k,\omega)
\ee
{It is not clear that the mean field approximation imposes an equally strong condition on $\delta(\vec k, \omega)$ as on $\Delta(\vec k)$, but we believe that a conservative estimate for it being a good approximation to treat $\tilde B$ as nearly uniform and weakly fluctuating 
is to require, }
\be{Deltaless}
\Delta(\vec k) \ \ \& \ \ \delta(\vec k) \ \  \ll 
(\phi_0)^2 \, .
\ee
Thus for large $\lambda$ (small $g$) all large $k$ fluctuations are effectively quenched.  The only remaining issues concern small $k$ fluctuations with $k\lambda \lesssim 1$.  

For an incompressible state, $\tilde \kappa(k,0) \sim k^2$, so again the fluctuations are automatically small.  In other words, for large but finite $\lambda$, the mean field theory should be accurate for all incompressible states, and only be questionable for compressible states at values of $k \lesssim 1/\lambda$.

\end{document}